\documentclass[12pt,epsfig,a4]{article}
\usepackage{graphicx}
\usepackage{amssymb}
\def\symbolfootnote[#1]#2{\begingroup%
\def\thefootnote{\fnsymbol{footnote}}\footnote[#1]{#2}\endgroup}

\begin{document}
\vspace*{-0.5in}
\begin{flushright}
\today\\
\end{flushright}
\vspace{0.1in}
\begin{center}
\begin{Large}
{\bf A Possible Future Long Baseline Neutrino and Nucleon
 Decay Experiment with a 100~kton Liquid Argon TPC at Okinoshima
 using the J-PARC Neutrino Facility
}
\vspace{5mm}
\end{Large}
{\large
A.Badertscher$^1$, T.Hasegawa$^2$, T.Kobayashi$^2$, A.Marchionni$^1$, A.Meregaglia$^1$\symbolfootnote[1]{\setlength{\baselineskip}{4mm}Now
at IPHC, Universit\'e Louis Pasteur, CNRS/IN2P3, Strasbourg, France.}, 
T.Maruyama$^{2,3}$, 
K.Nishikawa$^2$, and A.Rubbia$^1$\\
{\it (1) ETH Z\"urich, (2) KEK IPNS, (3) University of Tsukuba} \\
}
\vspace{10mm}
\end{center}

\renewcommand{\baselinestretch}{2}
\large
\normalsize

\setlength{\baselineskip}{5mm}
\setlength{\intextsep}{5mm}
\renewcommand{\arraystretch}{0.5}
\renewcommand{\thefootnote}{\fnsymbol{footnote}}

\begin{abstract}
In this paper, we consider the physics performance of 
a single far detector composed of 
a 100~kton next  generation Liquid Argon Time Projection Chamber (LAr TPC)
possibly located at shallow depth, coupled
to the J-PARC neutrino beam facility with a realistic 1.66~MW operation of the Main Ring.
The new far detector could be located in the region of Okinoshima islands 
(baseline  $L\sim 658$~km).
Our emphasis is based on the measurement of the $\theta_{13}$ and $\delta_{CP}$ parameters,
possibly following indications for a non-vanishing $\theta_{13}$ in T2K,
and relies on the opportunity offered by the LAr TPC to reconstruct the incoming
neutrino energy with high precision compared to other large detector
technologies. We mention other possible baselines like for example
J-PARC-Kamioka (baseline $L\sim 295$~km),
or J-PARC-Eastern Korean coast (baseline $L\sim 1025$~km). Such a detector would also
further explore the existence of proton decays.
\end{abstract}

\vspace*{0.5in}
\clearpage

\section{Introduction}

\setcounter{figure}{0}
\setcounter{table}{0}
\indent
The``Quest for the Origin of Matter Dominated Universe" is a long standing puzzle of our physical world
(see e.g.~\cite{cpmatter}).
Answers to this question might come from further exploration of
\begin{itemize}
\item the Lepton Sector CP Violation by precise testing of the neutrino oscillation processes
\begin{itemize}
\item measure precisely the $\delta_{CP}$ and the mixing angle $\theta_{13}$;
\item examine matter effect in neutrino oscillation process and possibly
conclude the mass hierarchy of neutrinos.
\end{itemize}
\item Proton Decay:
\begin{itemize}
\item Search for p $\rightarrow$ $\nu$ K$^+$ and p $\rightarrow$ e $\pi^0$
in the range $10^{34\div 35}$ years
\end{itemize}
\end{itemize}
with assuming non-equilibrium environment in the evolution of universe.

The primary motivation of the T2K experiment\footnote{\setlength{\baselineskip}{4mm}We call T2K
the approved Tokai-To-Kamioka experiment using Super-Kamiokande as far detector.
New investigations or performance upgrades beyond this approved phase will presumably
have a different name.}~\cite{Itow:2001ee} is 
to improve the sensitivity to the $\nu_{\mu} \rightarrow \nu_e$ conversion phenomenon 
in the atmospheric regime by about an order of magnitude in the mixing
angle $\sin^2 2\theta_{13}$ compared to the CHOOZ experimental limit~\cite{Apollonio:1999ae}.
The discovery of a non-vanishing  $\theta_{13}$ angle would
ascertain the $3\times 3$ nature of the lepton flavor mixing matrix.
The measurements in T2K might
indicate that the $\theta_{13}$ angle is in a region where the 
simultaneous determination of neutrino 
mass hierarchy and the CP violating phase becomes possible.

The J-PARC accelerator complex which includes
the 180~MeV LINAC, the 3~GeV Rapid Cycling Synchrotron (RCS) and the
30-50~GeV Main Ring Synchrotron (MR) is planned to be commissioned in 2008.
The J-PARC neutrino beam facility, under
construction for the T2K experiment, is foreseen to begin
operation in 2009.  
The final goal for the T2K experiment 
is to accumulate an integrated proton power on target of $0.75$~MW$\times5\times 10^7$ seconds.
Within a few years of run, critical information,
which will guide the future direction of the neutrino physics, 
will be obtained based on the data corresponding to about 
$1\div2$~MW $\times$ 10$^7$ seconds integrated proton power on target 
(roughly corresponding to a 3$\sigma$ discovery at 
sin$^2$2$\theta$$_{13}$$>$ 0.05 and 0.03, respectively)~\cite{NP08}. 

It is well documented in the literature (see e.g. Refs.~\cite{Burguet-Castell:2001ez,Minakata:2001qm,Barger:2001yr})
that the most challenging task for next generation 
long baseline experiments is to unfold the unknown oscillation 
parameters $\sin^22\theta_{13}$, $\delta_{CP}$ and mass hierarchy, $\rm sgn(\Delta m^2_{31})$,
and from the measurement of the energy dependence of the oscillation signal to
resolve the so-called problems of ``correlations'' and ``degeneracies''.

The most important experimental aspects here are the beam profile (e.g.
the ability to cover with sufficient statistics the $1^{st}$ maximum
of the oscillation, the $1^{st}$ minimum, and the $2^{nd}$
maximum), the visible energy resolution of the detector, with which
the neutrino energy can be reconstructed, and the spectrum
of the misidentified background (e.g. $\pi^0$ spectrum, etc.).

One possible approach, advocated in the literature~\cite{Marciano:2001tz,Diwan:2003bp},
is to locate a very massive coarse Water Cerenkov detector at very long baselines ($L\geq 2500$~km) from a 
new multi-GeV neutrino source and to detect more than one oscillation maximum. 
In this kind of configuration, although the oscillation peaks are well separated in energy,
an important issue is the ability to subtract the backgrounds, in particular
that coming from misidentified neutral current events with leading $\pi^0$'s which introduce new sources of systematic
errors. In order to alleviate this problem, a different
approach assuming two similar far Water Cerenkov detectors, one located at Kamioka ($L\sim 295$~km)
and the other in Korea ($L\sim 1000$~km)
at the same OA2.5$^\circ$ from the J-PARC neutrino
beam, was discussed in Ref.~\cite{Ishitsuka:2005qi}. The two detectors would see the same
sub-GeV neutrino beam (in fact the same as in T2K) but the different baselines
would allow to study $E/L$ regions corresponding to the 1st and 2nd oscillation maxima.

The approach of our paper is different. We consider the physics performance of 
a single 100~kton next  generation liquid Argon Time Projection Chamber (LAr TPC)
located  in the region of Okinoshima islands corresponding to a (relatively modest) baseline
$L\sim 658$~km. A detector in this location will automatically see the J-PARC
neutrino beam at an off-axis angle of $\sim 0.8^\circ$. As we show below, this configuration
allows to study both 1st and 2nd oscillation maximum peaks with good
statistics. We rely on the realistic opportunity offered by the increase of intensity of the J-PARC MR 
and on the higher precision of the LAr TPC  than other large detectors to separate
the two peaks. In addition, the $\pi^0$ background is expected to be highly suppressed
thanks to the fine granularity of the readout, hence the main irreducible background
will be the intrinsic $\nu_e$ component of the beam.

At this stage of our investigations, 
the emphasis is placed on the measurement of the $\theta_{13}$ and $\delta_{CP}$ parameters,
possibly following indications for a non-vanishing $\theta_{13}$ in T2K.
We also compare the possibility to determine the $\sin^22\theta_{13}$-$\delta_{CP}$
parameters with a single detector configuration 
at other possible baselines like for example
the T2K (Tokai-Kamioka, OA2.5$^\circ$, $L\sim 295$~km) 
or the Tokai-Eastern Korean coast ($L\sim 1025$~km)
with an off-axis OA1.0$^\circ$~\cite{Hagiwara:2006vn,Meregaglia:2008qr}. More
complete physics performance studies will be presented elsewhere.

\section{A possible next step beyond T2K}
If a significant $\nu_e$ signal were to be observed at T2K, an immediate
step forward to a next generation experiment aimed at the discovery of CP-violation
in the lepton sector would be recommended with
high priority.
We have conducted a case study for the discovery of lepton sector CP violation
based on the J-PARC MR power improvement scenario.
Naturally, next generation far neutrino detectors for lepton sector CP violation discovery 
will be very massive and large. As a consequence, 
the same detector will give us the rare and important opportunity to 
discover proton decay. 
Thus, we also discuss the proton decay discovery potential
with a huge underground detector.

Compared with the T2K experimental conditions, lepton sector CP violation discovery 
requires
\begin{itemize}
\item an improved neutrino beam condition (intensity increase, broader energy spectrum,
possibly re-optimization of the focusing optics, ...);
\item an improved far neutrino detector (by improvements we primarily mean increased
signal reconstruction efficiency, better background separation, better
energy resolution, ..., and not only its volume).
\end{itemize}
Detector improvements include
\begin{itemize}
\item detector technology;
\item its volume;
\item its baseline and off-axis angle with respect to the neutrino source.
\end{itemize}
As for the neutrino beam intensity improvement, a realistic first 
step power improvement scenario at 
J-PARC Accelerator Complex has been recently analyzed and
proposed (LINAC energy is recovered to be 400 MeV, h=1 operation at 
RCS and 1.92 seconds repetition cycle operation
 at MR).
As for the detector technology, we assume a 100 kton Liquid Argon 
Time Projection Chamber for our case study.

The effects of the CP phase $\delta_{CP}$ appear either
\begin{itemize}
\item as a difference between $\nu$ and $\bar{\nu}$ behaviors
(this method is sensitive to the $CP$-odd term which vanishes
for $\delta_{CP}=0$ or 180$^\circ$);
\item in the energy spectrum shape of the appearance oscillated $\nu_e$ charged current (CC) events 
(sensitive to all the non-vanishing $\delta_{CP}$ values including 180$^\circ$).
\end{itemize}
Since antineutrino beam conditions are known to be more difficult than those
for neutrinos (lower beam flux due to leading charge effect in proton collisions
on target, small antineutrinos cross-section at low energy, etc.), we
concentrate on the possibility to precisely measure the 
$\nu_e$ CC appearance energy spectrum shape with high resolution
during a neutrino-only run.

\section{The liquid Argon Time Projection Chamber}
The liquid Argon Time Projection Chamber (LAr TPC) (See Ref.~\cite{t600paper}
and references therein) is
a powerful detector for uniform and high accuracy imaging of massive active volumes. 
It is based on the fact that in highly pure Argon, ionization tracks can be drifted
over distances of the order of meters. 
Imaging is provided by position-segmented electrodes at the end of the drift path, continuously recording the 
signals induced. $T_0$ is provided by the prompt scintillation light. 

In this paper, we assume a LAr TPC detector with a mass of the order of 100~kton, for example of a kind 
based on the GLACIER concept~\cite{Rubbia:2004tz}. Unlike other liquid Argon TPCs
operated or planned which rely on immersed wire chambers to readout the ionization
signals, a double phase operation with charge extraction and amplification in the vapor
phase is considered here in order to allow for very long drift paths and for improved 
signal-to-noise ratio. We assume that successful application of such novel methods
will be an important milestone in the R\&D for very large LAr TPC detectors in the range
of 100~kton. At this stage, a ton-scale prototype based on
this scheme has been developed and is under test~\cite{Rubbia:2005ge}.
Challenges to realize liquid Argon TPC's with a scale relevant to this paper
have been reviewed e.g. in Ref.~\cite{NP08lar}. 

We refer to previous
physics performance studies assuming such a detector configuration~\cite{Meregaglia:2008qr,Meregaglia:2006du} 
and recall that since Liquid Argon TPC has advantages on
\begin{itemize}
\item good energy resolution/reconstruction,
\item good background suppression,
\item good signal efficiency
\end{itemize}
it is suitable for a precision measurement of the neutrino
energy spectrum to extract CP information. Thus we concentrate on 
the $\nu_e$ appearance energy spectrum shape measurement to
extract leptonic CP phase information.
%

%

\section{The choice of far location: Okinoshima}
\begin{figure}[htb]
\begin{center}
\includegraphics[angle=0, width=0.9\textwidth]{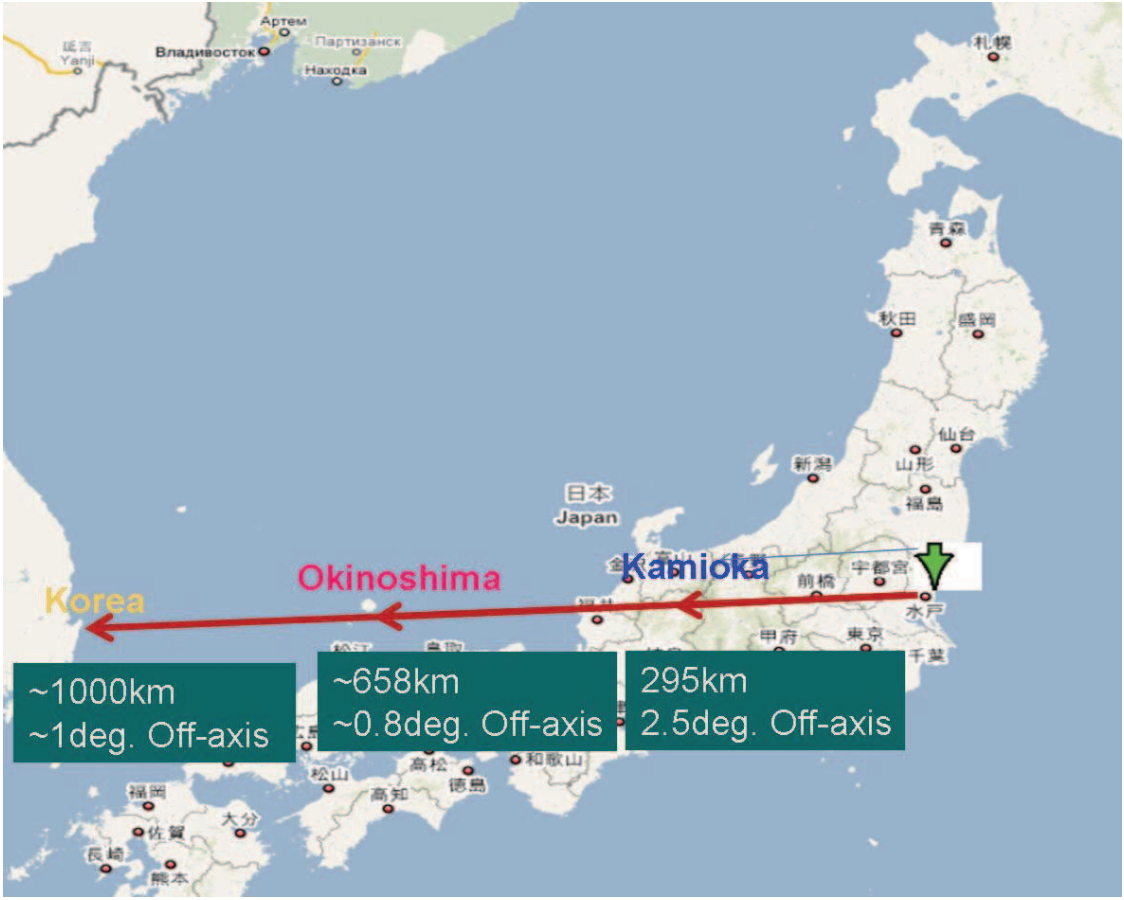}
\caption{\setlength{\baselineskip}{4mm}
  Possible baseline candidates for 100 kton LAr TPC detector,
  Kamioka (short-baseline), Okinoshima (middle-baseline) 
  and Korea (long-baseline).  
}
\label{Fig:BLcand}
\end{center}
\end{figure}

\indent
The J-PARC neutrino beam line was designed and constructed in order to allow an off-axis angle with respect
to Super-Kamiokande (SK) between 2.5$^\circ$ and 3$^\circ$. A beam setup yielding an OA2.5$^\circ$ at SK was
chosen for the T2K experiment. 
In this configuration, the center of the T2K neutrino beam will go through underground beneath SK, and 
will automatically reach the Okinoshima island region with an off-axis angle $\sim 0.8^\circ$ and eventually the sea level east of the Korean shore
with an off-axis  angle $\sim 1^\circ$. Larger off-axis angles are obtained moving inland Korea (either north, south or west). 
Figure~\ref{Fig:BLcand} shows these baseline options using the same
beam configuration as the T2K experiment. Parameters for different baselines 
and beam axes are summarized in Figure~\ref{Fig:BLcand} and 
Table~\ref{Tab:BLcand}. 

\begin{table}[ht]
\begin{center}
\begin{tabular}{lccc} 
\hline \hline 
\multicolumn{4}{c}{Same beam configuration as T2K experiment}\\
\hline
   & Kamioka & Okinoshima & Korea eastern shore \\ \hline
  Baseline & 295 km & $\sim$658 km & $\sim$1000 km \\
  Off-axis angle & 2.5$^{\circ}$ & $\sim$0.8$^{\circ}$ & $\sim$1.0$^{\circ}$  \\
\hline
\end{tabular}
\caption{\setlength{\baselineskip}{4mm}
Parameters for baseline and axis with respect to beam of the 
candidates are summarized.
}
\label{Tab:BLcand}
\end{center}
\end{table}

The neutrino flavor oscillation probability including atmospheric, solar and interference
terms, as well as matter effects, can expressed using the following 
equation~\cite{Freund:2001pn,Cervera:2000kp,Hagiwara:2006vn}
\begin{equation}
P(\nu_{e} \rightarrow \nu_{\mu}) \sim \sin^{2} 2\theta_{13} \cdot T_{1} + \alpha \cdot \sin \theta_{13} \cdot (T_{2}+T_{3}) + \alpha^{2} \cdot T_{4} .
\end{equation}
where, 
\begin{eqnarray}
T_{1} &=& \sin^2 \theta_{23} \cdot \frac{\sin^2 [(1-A) \cdot \Delta]}{(1-A)^{2}} \nonumber\\
T_{2} &=& \sin \delta_{CP} \cdot \sin 2\theta_{12} \cdot \sin 2\theta_{23} \cdot \sin \Delta \frac{\sin(A\Delta)}{A} \cdot \frac{\sin[(1-A)\Delta]}{(1-A)}  \nonumber\\
T_{3} &=& \cos \delta_{CP} \cdot \sin 2\theta_{12} \cdot \sin 2\theta_{23} \cdot \cos \Delta \frac{\sin(A\Delta)}{A} \cdot \frac{\sin[(1-A)\Delta]}{(1-A)}  \nonumber\\
T_{4} &=& \cos^2 \theta_{23} \cdot \sin^2 2\theta_{12} \frac{\sin^2(A\Delta)}{A^2}.
\end{eqnarray}
where 
$\alpha \equiv \frac{\Delta m_{21}^2}{\Delta m_{31}^2}$, $\Delta \equiv
\frac{\Delta m_{31}^2 L}{4E}$, $A \equiv \frac{2\sqrt{2} G_{F} n_{e}
E}{\Delta m_{31}^2}$. $\Delta m_{31} = m_3^2 - m_1^2$, 
$\Delta m_{21} = m_2^2 - m_1^2$, $\theta_{13}$ is the mixing angle of
the 1st and 3rd generations, while $\theta_{12}$ is that for 1st and 2nd,  
and $\theta_{23}$ is that for 2nd and 3rd generations.

\begin{figure}[ht]
\begin{center}
\includegraphics[angle=0, width=0.9\textwidth]{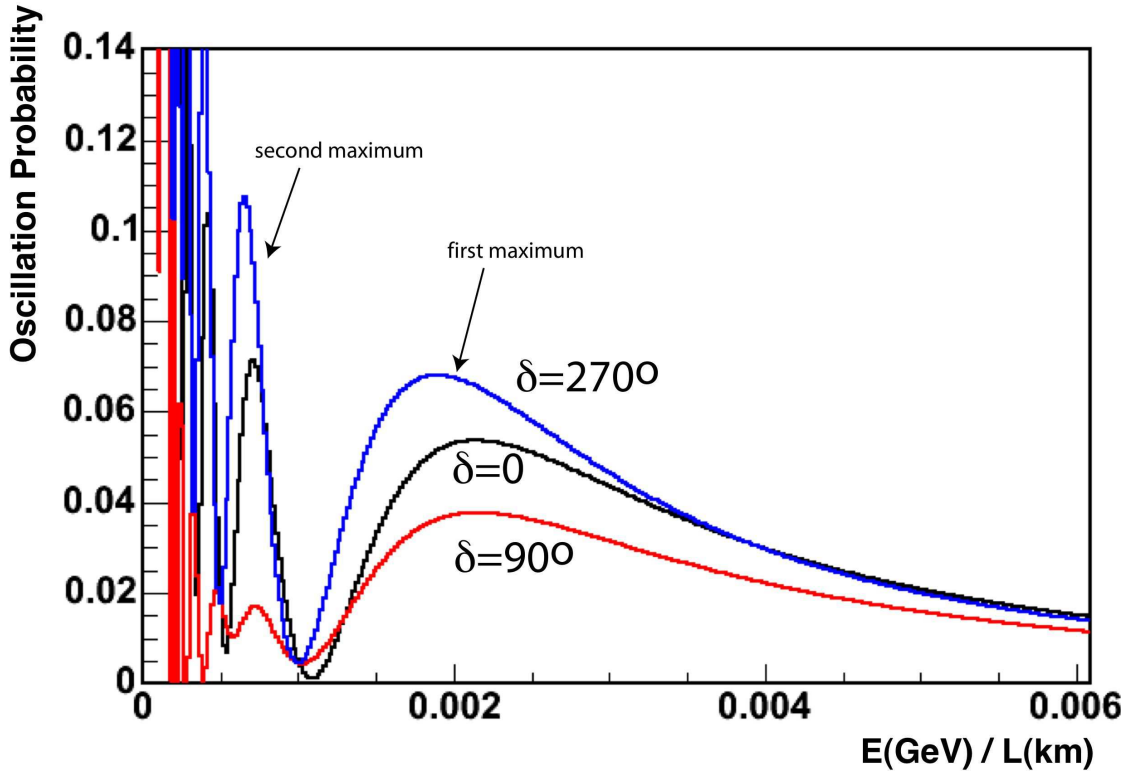}
\caption{\setlength{\baselineskip}{4mm}
Probability for $\nu_\mu\rightarrow \nu_e$ oscillations as a function of the E(GeV)/L(km)
for $\sin^2 2\theta_{13}=0.1$ (for other oscillation parameters see text). Black curve shows 
the case for the $\delta_{CP}$=0, red shows that for
 $\delta_{CP}$=90$^\circ$, blue shows that for $\delta_{CP}$=270$^\circ$, respectively. 
}
\label{Fig:OPloe}
\end{center}
\end{figure}

The term which includes $T_1$ is the ``atmospheric term'', those
including $T_2$ and $T_3$ are ``interference terms'', and 
the one that includes $T_4$ is the ``solar term''.
The ``interference terms'' are sensitive to the $\delta_{CP}$ phase and
play therefore an important role to extract the CP phase.

For definite calculations, we use the following parameters (we assume that most of these
parameters will be precisely measured within the timescale of the one discussed
here):
\begin{itemize}
\item $\theta_{23}$ is $\pi$/4.
\item $\theta_{12}$ is 0.572904 rad. 
\item $\Delta m^2_{21}$ is 8.2$\times$10$^{-5}$ eV$^{2}$. 
\item $|\Delta m^2_{31}|$ is 2.5$\times$10$^{-3}$ eV$^{2}$.
\item Earth density for matter effects are 2.8 g/cm$^{3}$.
\item normal hierarchy is assumed unless mentioned otherwise.
\end{itemize}

Figure~\ref{Fig:OPloe} shows the oscillation probability as a function 
of the E(GeV)/ L(km) (neglecting for the moment matter effects). 
In the plot, the mixing angle $\sin^2 2\theta_{13}$ was assumed to be 0.1.
If the distance between source and detector is fixed, the curves can be easily translated to
that for the expected neutrino energy spectrum of the oscillated events.
As can be seen, if the neutrino energy spectrum of the oscillated events could be reconstructed with sufficiently 
good resolution in order to distinguish first and second maximum, useful information to extract the 
CP phase would be available even only with a neutrino run.

We note that the following observables depend on leptonic CP phase:
\begin{itemize}
\item position/height of the first oscillation maximum peak;
\item position/height of the second oscillation maximum peak.
\end{itemize}
The position of the first oscillation minimum is unaffected by it
(in this point all terms of the oscillation probability vanish except the solar term).
To effectively experimentally study these observables, we point out that:
\begin{itemize}
\item the second oscillation maximum peak should end up at
sufficiently high energy in order to be measurable;
\item the beam should have a sufficiently wide energy range to
cover the 1st and 2nd maximum;
\item the neutrino flux should be maximized in order to increase as much as possible the statistical significance
of the first and second maxima. 
\end{itemize}
In order to cover a wider energy range,  we accordingly favor a detector location which is near on-axis.
If one assumes that the second oscillation maximum has to be located at an energy larger than  
about 400 MeV, the oscillation baseline should be longer than  about 600~km.
In addition, in order to collect enough statistics, we choose a baseline which is not too much 
longer than above stated.

Taking into account all of the above mentioned considerations, we privilege the Okinoshima region:
placing a detector in an appropriate location on the island~\footnote{\setlength{\baselineskip}{4mm}The exact location of 
the potential experiment in Okinoshima has not
yet been investigated, but we note that the island has several hills.} will 
probe neutrino oscillations at a baseline of $\sim 658$~km away from the source
at an off-axis angle of $\sim 0.8^\circ$.

\section{Neutrino Flux and expected Event Rates}
\indent

We assume realistic parameters of the J-PARC beam after all accelerator complex upgrades described in
the KEK road-map are accomplished, as follows~\cite{NP08}:
\begin{itemize}
\item The average beam power is reaching 1.66 MW;
\item A total of 3.45 $\times$ 10$^{21}$ POT is delivered on target per year;
\item The optimal kinetic energy of the incident protons is 30 GeV.
\end{itemize}

To remain conservative, we focus on an analysis which uses a neutrino run only
during five years under the best J-PARC beam assumption. An antineutrino beam
(opposite horn polarity) might be considered in a second stage in order to cross-check
the results obtained with the neutrino run (in particular see Section~\ref{sec:masshier} on the 
mass hierarchy problem). The
parameters of the Okinoshima location and the assumed beam are therefore as follows:
\begin{itemize}
\item Distance from J-PARC is 658 km;
\item The axis of the beam is off by 0.76$^\circ$;
\item five years operation with horn setting to neutrino run.
\end{itemize}

\begin{figure}[htb]

\begin{minipage}[b]{0.5\linewidth}

\centering
\includegraphics[angle=0, height=8cm]{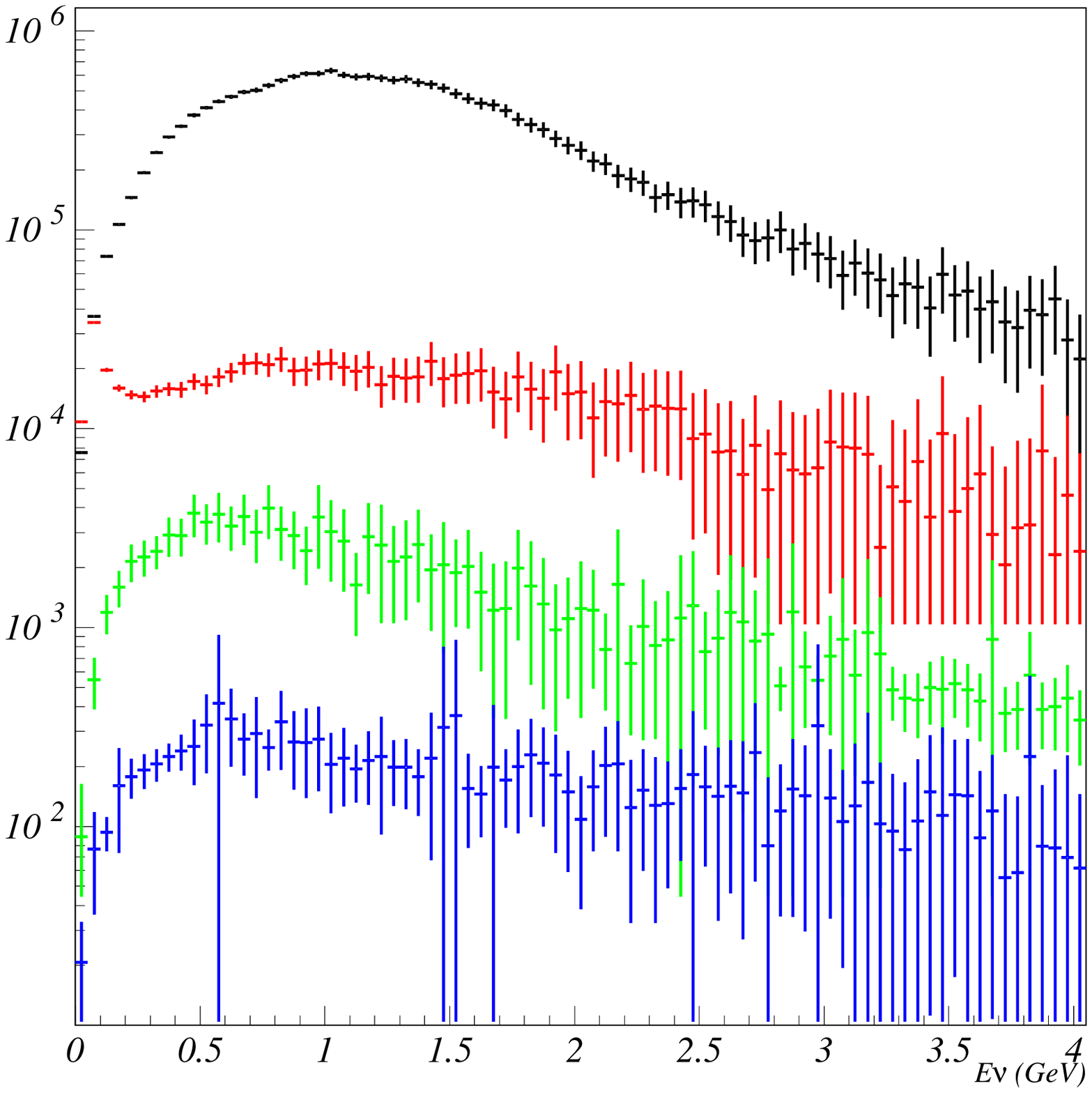}
\caption{\setlength{\baselineskip}{4mm}
Calculated neutrino flux under the assumptions (3.45$\times 10^{21}$
 POT). Black shows $\nu_{\mu}$, red shows $\bar{\nu_{\mu}}$, 
green shows $\nu_e$, blue shows $\bar{\nu_{e}}$ fluxes, 
respectively. 
}
\label{Fig:flux}

\end{minipage}
\hspace{0.5cm}
\begin{minipage}[b]{0.5\linewidth}

\centering
\includegraphics[angle=0, height=8cm]{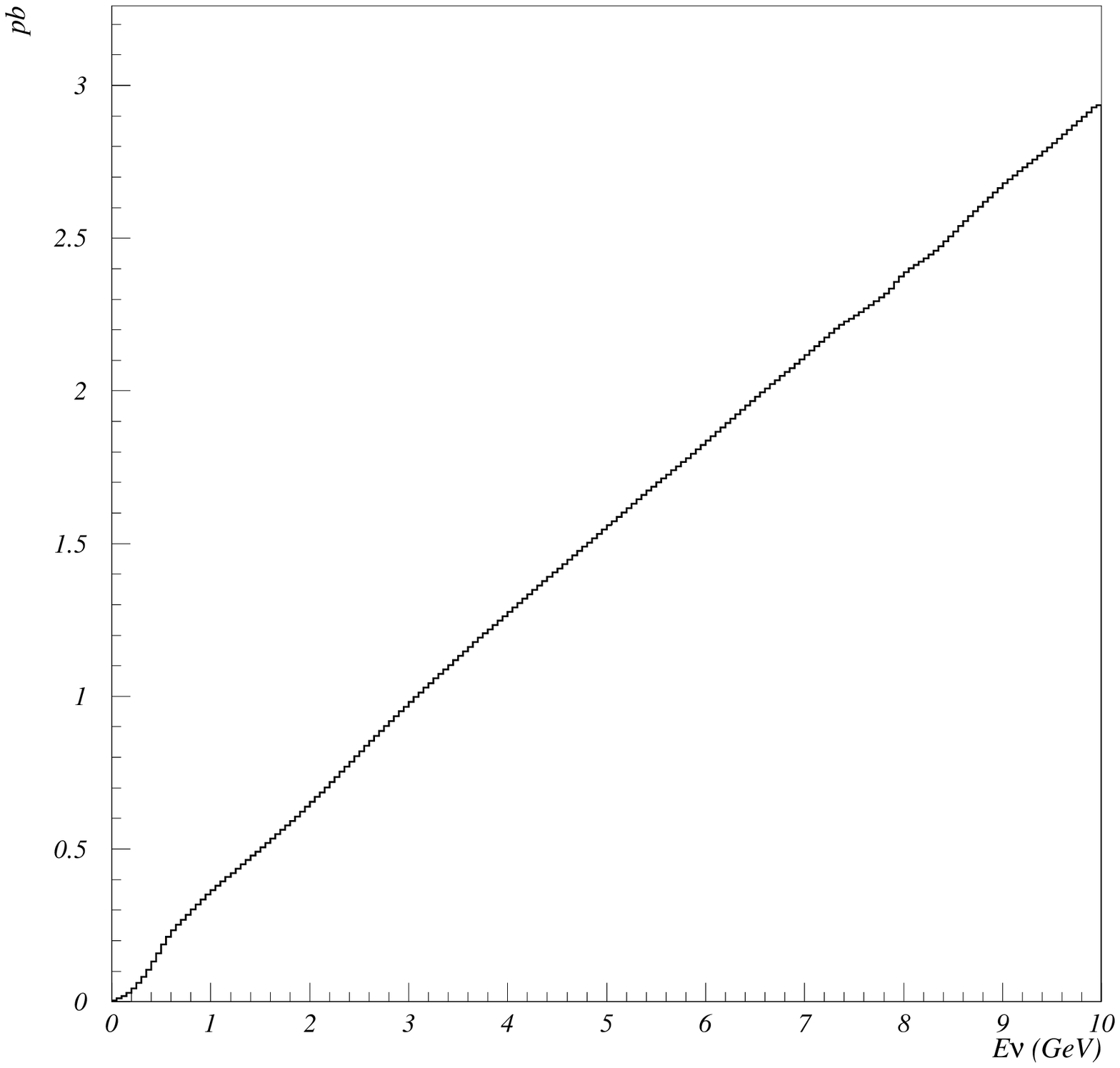}
\caption{\setlength{\baselineskip}{4mm}
Calculated neutrino-Argon cross section, unit by pb.
Nuclear effects are taken into account.
}
\label{Fig:XS}

\end{minipage}
\end{figure}

The expected neutrino flux calculated under these assumptions is shown in 
Figure~\ref{Fig:flux} where the curves correspond to one year run 
(3.45$\times 10^{21}$ POT). The black, red, green, blue lines show $\nu_{\mu}$,
$\bar\nu_{\mu}$, $\nu_e$, $\bar\nu_{e}$ fluxes, respectively.

The interacting neutrino cross section on Argon was computed using the NUANCE programme~\cite{Ref:NUANCE}.
We use the followings parameters for the cross section calculation:
\begin{itemize}
\item Number of protons is 18, and that of neutrons is 22.
\item Medium density is 1.4 g/cm$^{3}$.
\item Nucleus Fermi Gas model with binding energy of 30~MeV and Fermi momentum of 242~MeV/c.
\end{itemize}

Taking these into account, we obtain the neutrino cross section shown in Figure~\ref{Fig:XS}.
Table~\ref{Tab:NuRate} shows that total number of charged current (CC)
events at Okinoshima for the null oscillation case and the number of CC $\nu_e$ events
from $\nu_\mu\rightarrow \nu_e$ oscillations for three different mixing angles, and
 in various $\delta_{CP}$ scenarios, normalized to five years neutrino run
at 1.66~MW and 100~kton fiducial mass.

\begin{table}[ht]
\begin{center}
\begin{tabular}{lcccc} 
\hline \hline 
\multicolumn{5}{c}{Events for 100~kton at Okinoshima normalized to 5 years at 1.66~MW beam power}\\
\hline
 & $\nu_{\mu}$ & $\nu_e$ & $\bar{\nu}_{\mu}$ & $\bar{\nu}_e$\\ \hline
Beam components (null oscillation)  &  82000  & 750 & 1460 & 35 \\
\hline \hline 
\multicolumn{5}{c}{$\nu_\mu\rightarrow \nu_e$ oscillations }\\
  $\delta_{CP}=$& $0^{\circ}$ & 90$^{\circ}$ & 180$^{\circ}$ & 270$^{\circ}$\\ \hline
  $\sin^2 2\theta_{13}$ = 0.1 & 2867 & 2062 & 2659 & 3464  \\
  $\sin^2 2\theta_{13}$ = 0.05 & 1489 & 1119 & 1342 & 1908  \\
  $\sin^2 2\theta_{13}$ = 0.03 & 942 & 506 & 829 & 1266  \\
\hline
\end{tabular}
\caption{\setlength{\baselineskip}{4mm}
Number of CC events at Okinoshima: beam components for null oscillation,
and oscillated events in various $\delta_{CP}$ scenarios
(normalized to five years neutrino run)
}
\label{Tab:NuRate}
\end{center}
\end{table}

The number of expected events is already a good indicator of the signal, but for this
analysis we use binned likelihood fitting for the energy spectrum, 
therefore all information including total number of events is 
taken into account, as described in the next section.

\section{Expected Energy Spectrum of $\nu_{e}$ CC events}
\indent

Images taken with a liquid Argon TPC are comparable with pictures from bubble 
chambers. As it is the case in bubble chambers, events can be analyzed by reconstructing 
3D-tracks and particle types for each track in the event image, with a lower energy 
threshold of few MeV for electrons and few tens of MeV for protons. The particle type 
can be determined from measuring the energy loss along the track ($dE/dx$) or from 
topology (i.e. observing the decay products). Additionally, the electronic readout allows
to consider the volume as a calorimeter adding up all the collected ionization charge.
The calorimetric performance can be excellent, as we will show in Section~\ref{sec:eneres},
depending on event energy and topology.

In order to understand the effect of resolution on physics performance,
we show in this section the $\nu_e$ charged current (CC) event energy spectra using 
different simple models. One is the perfect resolution case as a
reference, and others are 100 MeV/200 MeV Gaussian resolution cases for
more realistic cases. 
This ``resolution'' should include 
neutrino interaction effects as well as detector resolution.
Possible smearing or backgrounds affecting the measurement of the
neutrino energy spectrum are listed below subdivided into 4 classes:

\begin{itemize}
\item Neutrino interaction:
      \begin{itemize}
       \item Fermi motion and nuclear binding energy,
       \item Nuclear interactions of final state particles within the hit nucleus (FSI),
       \item Vertex nuclear remnant effects (e.g. nuclear break-up signal),
       \item Neutral Current (NC) $\pi^{0}$ event shape including vertex activity.           
      \end{itemize}
\item Detector medium:
      \begin{itemize}
       \item Ionization processes,
       \item Scintillation processes,
       \item Correlation of between amount of charge and light,
       \item Charge and light quenching,
       \item Hadron transport in Argon and secondary interactions,
       \item Charge diffusion and attenuation due to impurity attachment.
      \end{itemize}      
\item Readout system including electronics system:
      \begin{itemize}
       \item signal amplification or lack thereof,
       \item signal-to-noise ratio,
       \item signal shaping and feature extraction.
      \end{itemize}       
\item Reconstruction:
      \begin{itemize}
       \item Pattern recognition
       \item Background processes (NC $\pi^0$, $\nu_\mu$~CC, ...) and their event shape
       \item Particle identification efficiency and purity
      \end{itemize}       
\end{itemize}

Very few real events in liquid Argon TPCs have so far been accumulated to allow a full understanding
of these complex effects and their interplay. The only sample comes from 
a small 50~lt chamber developed by the ICARUS Collab.
and exposed to the CERN WANF high energy neutrino beam which collected
less than 100~quasi-elastic events~\cite{Arneodo:2006ug}.

It is clear that the energy resolution will depend on several detector
parameters, including the readout pitch, the readout method chosen and
on the resulting signal-over-noise ratio ultimately affecting the reconstruction of the events.
We therefore stress that significantly
improved experimental studies with prototypes exposed to
neutrino beams of the relevant energies and sufficient statistics 
are mandatory to assess and understand these effects.
In the meantime, we give in Section~\ref{sec:eneres} preliminary
estimates for potentially achievable energy resolutions in a liquid Argon medium. 
The results are based on full GEANT3~\cite{GEANT3} simulation of the energy deposited by final
state particles in the detector volume, however do not include all
possible contributing effects.

\begin{figure}[htb]

\begin{minipage}[b]{0.5\linewidth}

\centering
\includegraphics[angle=0, width=1.05\textwidth]{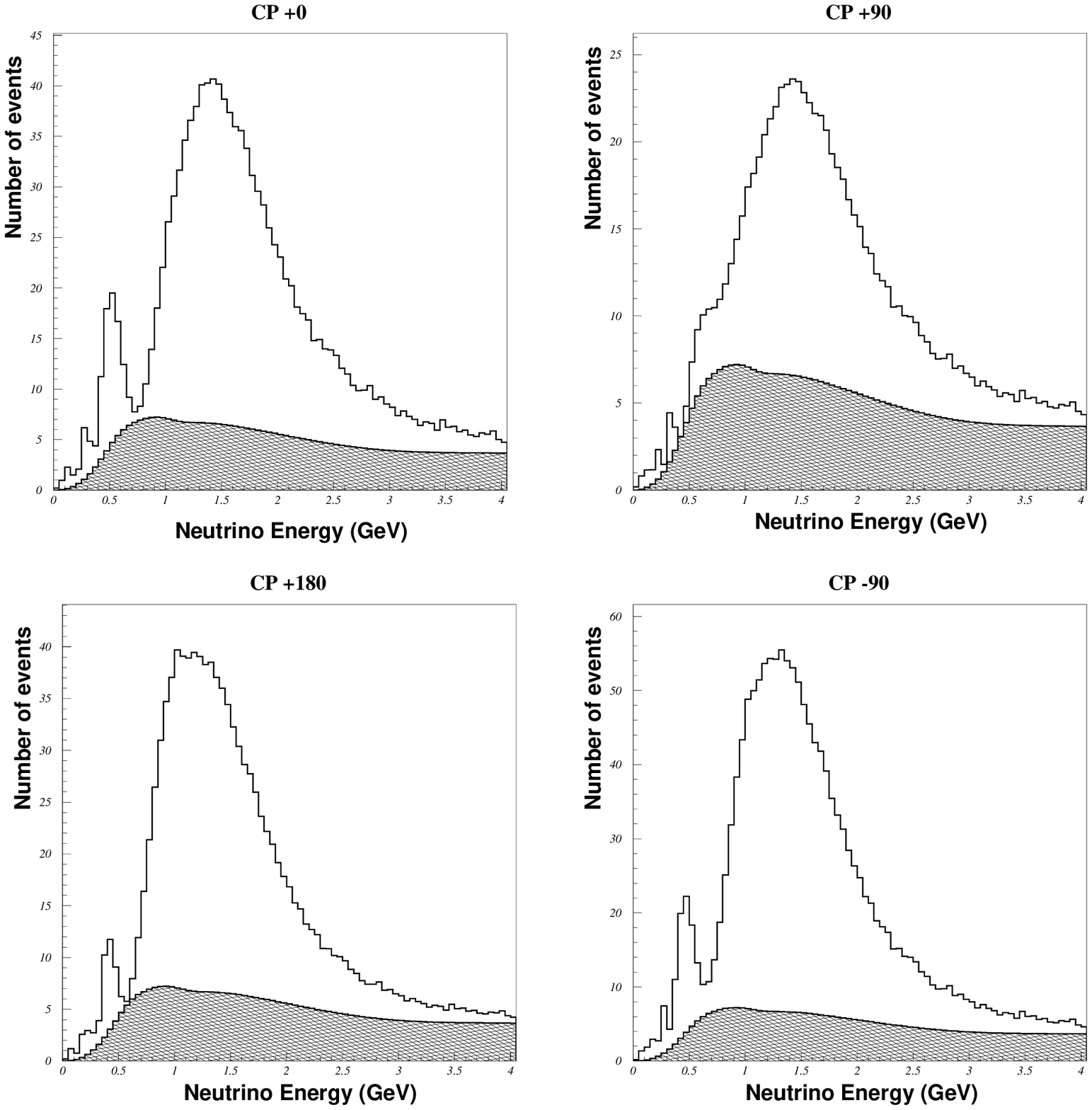}
\caption{\setlength{\baselineskip}{4mm}
Energy spectra at $\sin^2 2\theta_{13}$=0.03 case, but 
 $\delta_{CP}$ = 0$^{\circ}$ (top-left), 90$^{\circ}$ (right-top),
180$^{\circ}$ (left-bottom), 270$^{\circ}$ (right-bottom) cases. 
}
\label{Fig:ES0mev}

\end{minipage}
\hspace{0.5cm}
\begin{minipage}[b]{0.5\linewidth}

\centering
\includegraphics[angle=0, width=\textwidth]{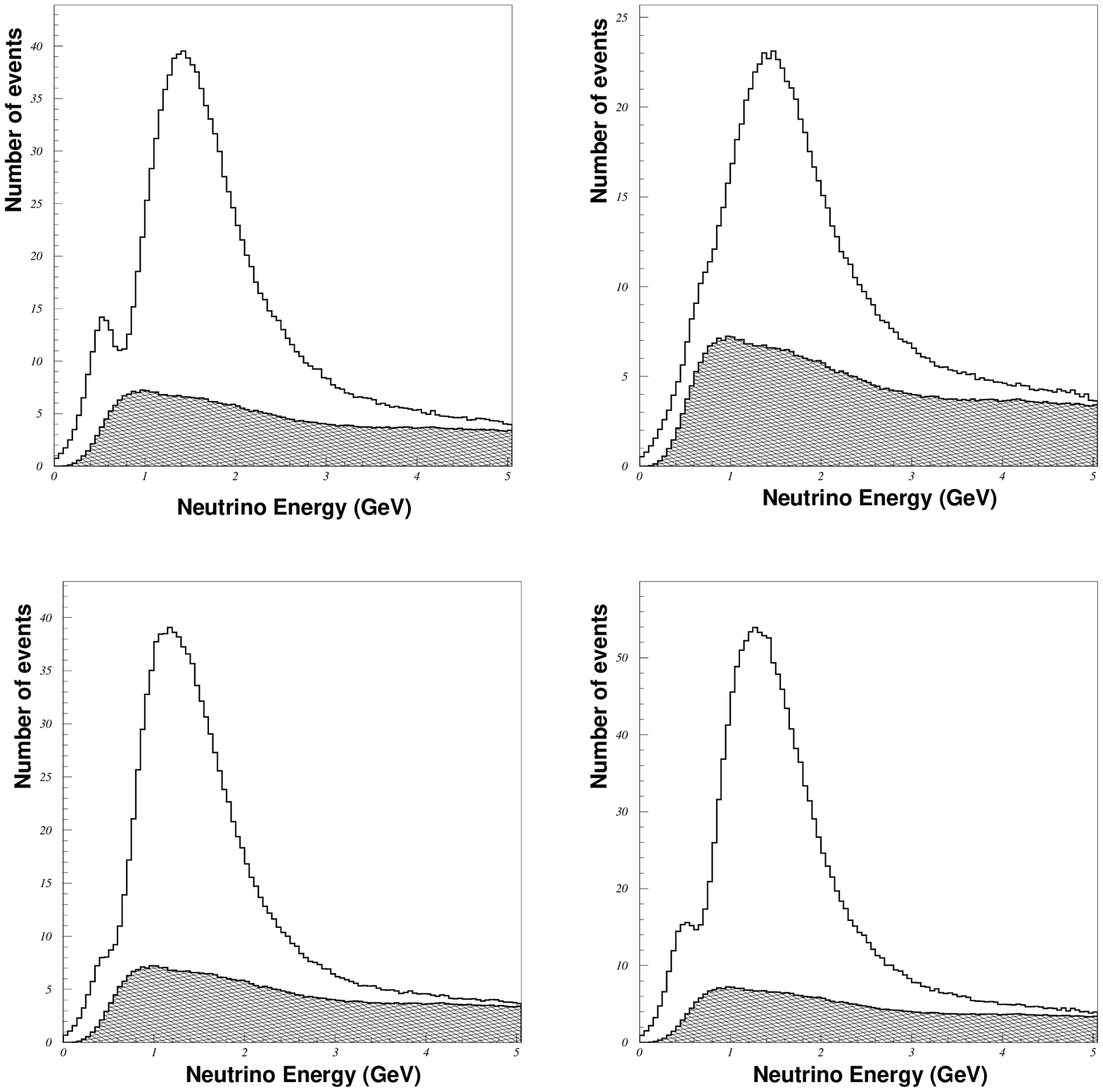}
\caption{\setlength{\baselineskip}{4mm}
Energy spectra assuming Gaussian 100 MeV smearing. 
 $\delta_{CP}$ = 0$^{\circ}$ (top-left), 90$^{\circ}$ (right-top),
180$^{\circ}$ (left-bottom), 270$^{\circ}$ (right-bottom) cases. 
}
\label{Fig:ES100mev}

\end{minipage}

\end{figure}

Figure~\ref{Fig:ES0mev} shows the energy spectra of electron neutrino 
at the cases 
of $\delta_{CP}$ equal 0$^{\circ}$, 90$^{\circ}$, 180$^{\circ}$, 
270$^{\circ}$, respectively. Shaded region is common for all 
plots and it shows the background from beam $\nu_e$.
Here perfect resolution is assumed for reference to later cases.

\begin{figure}[htb]
\begin{center}
\includegraphics[angle=0, width=9.0cm]{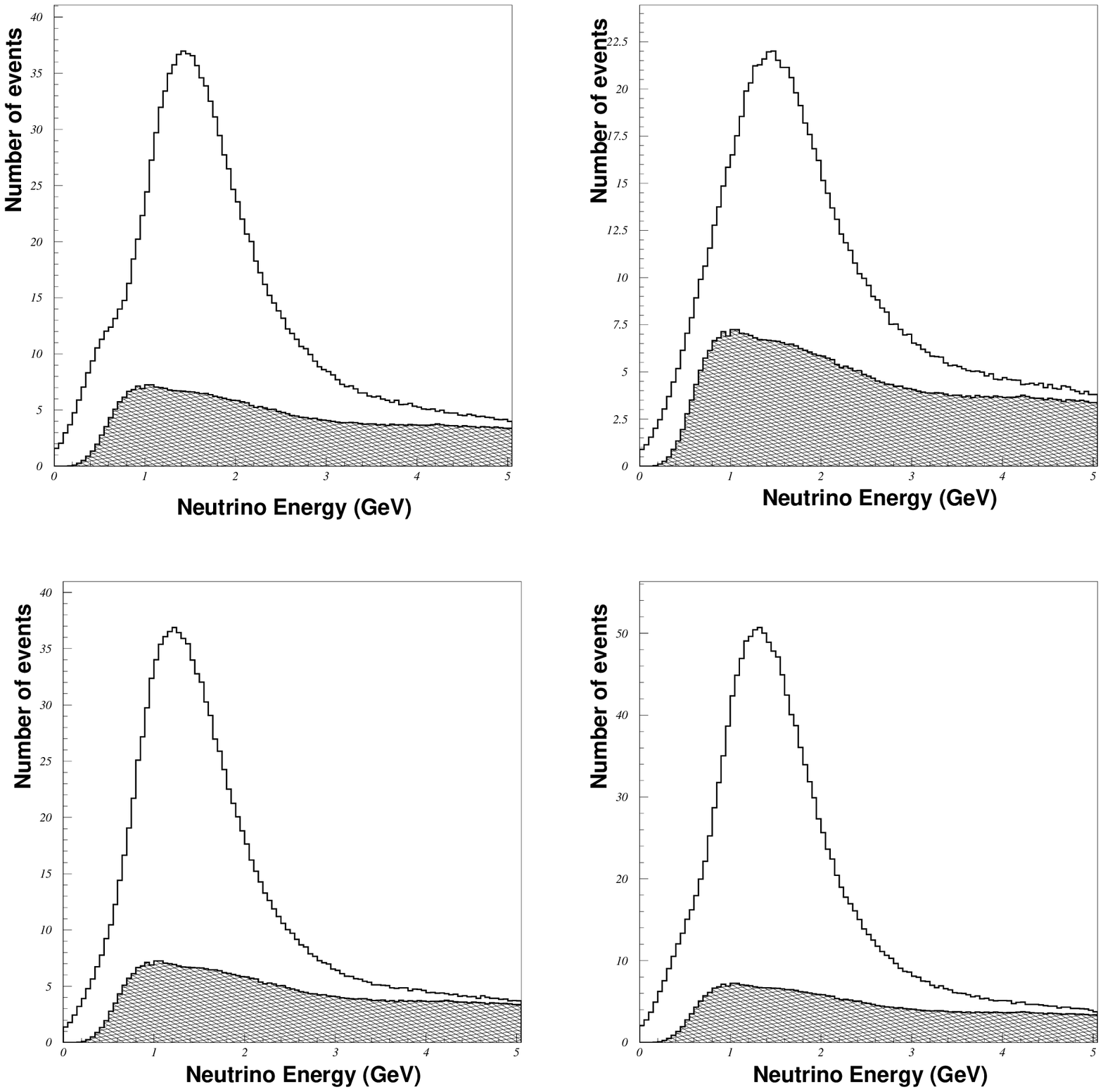}
\caption{\setlength{\baselineskip}{4mm}
Energy spectra assuming Gaussian 200 MeV smearing. 
 $\delta_{CP}$ = 0$^{\circ}$ (top-left), 90$^{\circ}$ (right-top),
180$^{\circ}$ (left-bottom), 270$^{\circ}$ (right-bottom) cases. 
}
\label{Fig:ES200mev}
\end{center}
\end{figure}

If we smear the energy spectra shown in Figure~\ref{Fig:ES0mev}
with Gaussian of sigma equals to 100 or 200 MeV independent from
original neutrino energy, we obtain spectra shown in   
Figures \ref{Fig:ES100mev} and \ref{Fig:ES200mev}. 
As seen easily, an energy resolution below 100~MeV is crucial since
the robustness of the neutrino oscillation is directly determined by the
visible second oscillation peak around 400 MeV in the energy spectrum.
In 200 MeV resolution case, the second peak is hidden by the smearing
of the 1st oscillation maximum peak.

\section{\setlength{\baselineskip}{4mm}
Oscillation parameters measurement from energy spectrum analyses}
\indent


Assuming all others were measured very precisely, 
there are only two free parameters $\sin^2 2\theta_{13}$ and $\delta_{CP}$
to be fitted using the energy spectra.
As shown in Figure~\ref{Fig:OPloe}, the value of $\delta_{CP}$ varies the energy 
spectrum, especially the first and the second oscillation peaks (heights
and positions), therefore comparison of the peaks
determine the value $\delta_{CP}$, while the value of $\sin^2 2\theta_{13}$ changes 
number of events predominantly.

\begin{figure}[htbp]
\begin{center}
\includegraphics[angle=0, width=8.0cm]{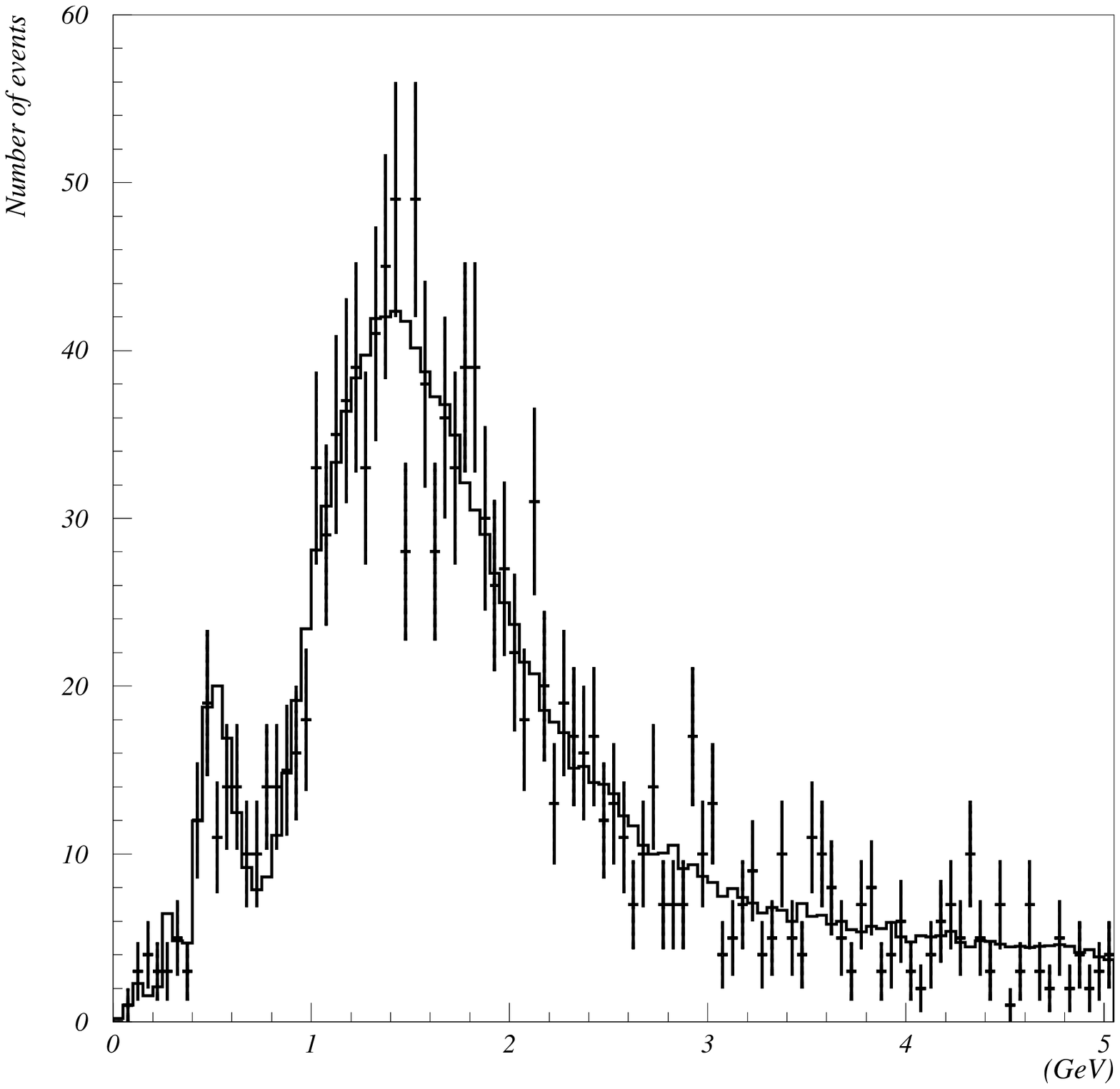}
\caption{\setlength{\baselineskip}{4mm}
 Typical one pseudo-experiment to show the validity of the fitting.
 Histogram shows the best-fit (best oscillation shape + background), and
 the crosses show the pseudo-data. 
}
\label{Fig:aPE}
\end{center}
\end{figure}

To fit the free parameters, a binned likelihood method is used.  
The Poisson bin-by-bin probability of the observed data from the expected events 
(for an assumed pair of values ($\sin^2 2\theta_{13}$, $\delta_{CP}$))
 is calculated. The fit procedure is  validated by testing the result
 on a pseudo-experiment. Figure~\ref{Fig:aPE} shows one typical pseudo-experiment for the perfect
resolution case (with $\delta_{CP}$ equals to 0, and $\sin^2 2\theta_{13}$
equals to 0.1). Best fit gives reasonable result.

At this stage, we only take statistical uncertainty into account for the fitting, thus
other systematic uncertainties like far/near ratio, beam $\nu_{e}$ shape,
energy scale, and so forth are not considered.
Also, oscillated signal and beam $\nu_{e}$ are only accounted 
in the fit, i.e. other background 
like neutral current $\pi^{0}$ or beam $\bar\nu_{e}$ contamination
is assumed to be negligible compared to beam $\nu_{e}$ contamination.

\begin{figure}[htb]

\begin{minipage}[b]{0.5\linewidth}

\centering
\includegraphics[angle=0, width=9.0cm]{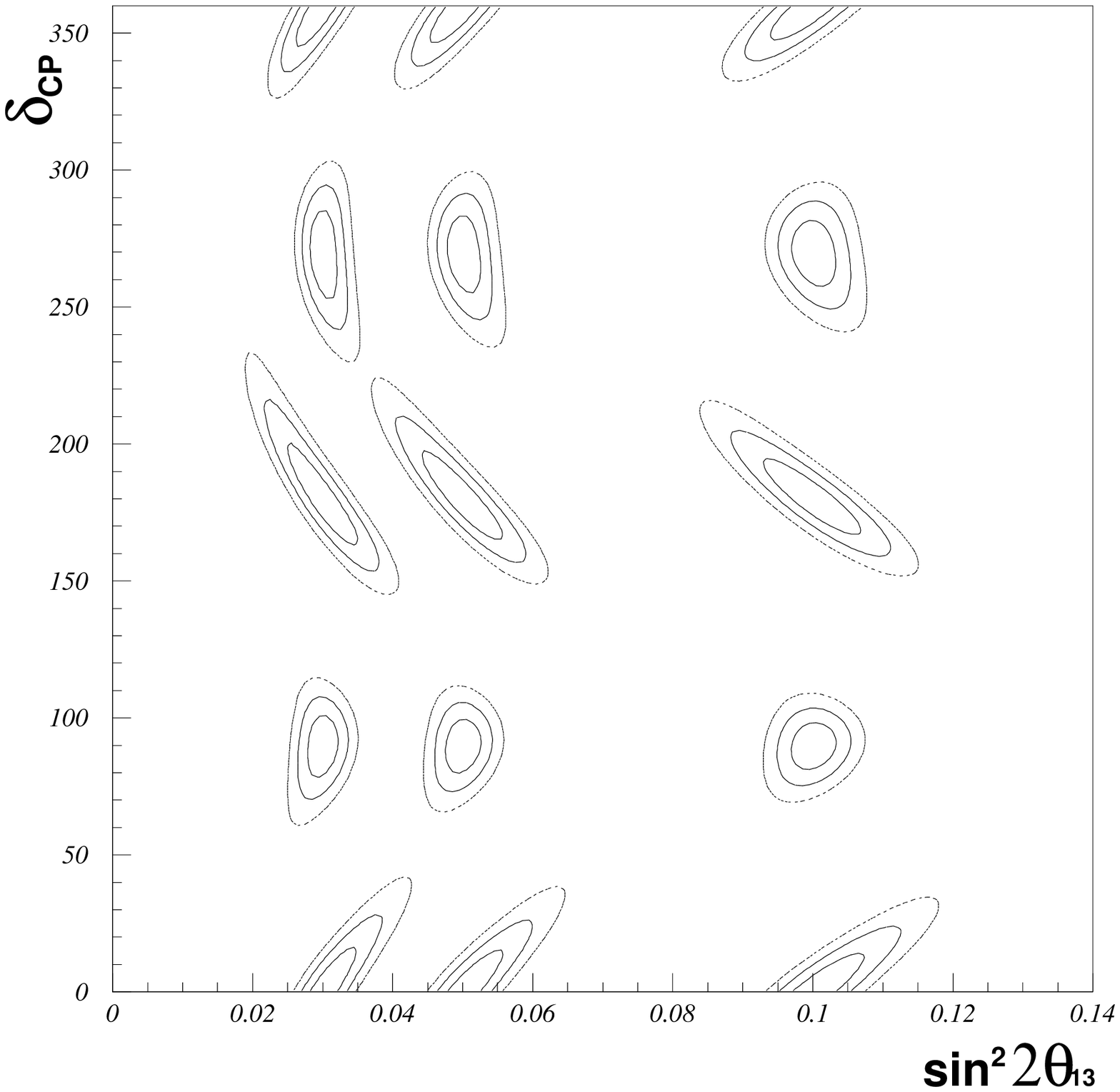}
\caption{\setlength{\baselineskip}{4mm}
 Allowed regions in the perfect resolution case.
 Twelve allowed regions are overlaid for twelve true values,
 $\sin^2 2\theta_{13}$=0.1, 0.05, 0.02, and $\delta_{CP}$=0$^{\circ}$,
 90$^{\circ}$, 180$^{\circ}$, 270$^{\circ}$, respectively.
}
\label{Fig:Allowed_0mev}

\end{minipage}
\hspace{0.5cm}
\begin{minipage}[b]{0.5\linewidth}

\centering
\includegraphics[angle=0, width=9.0cm]{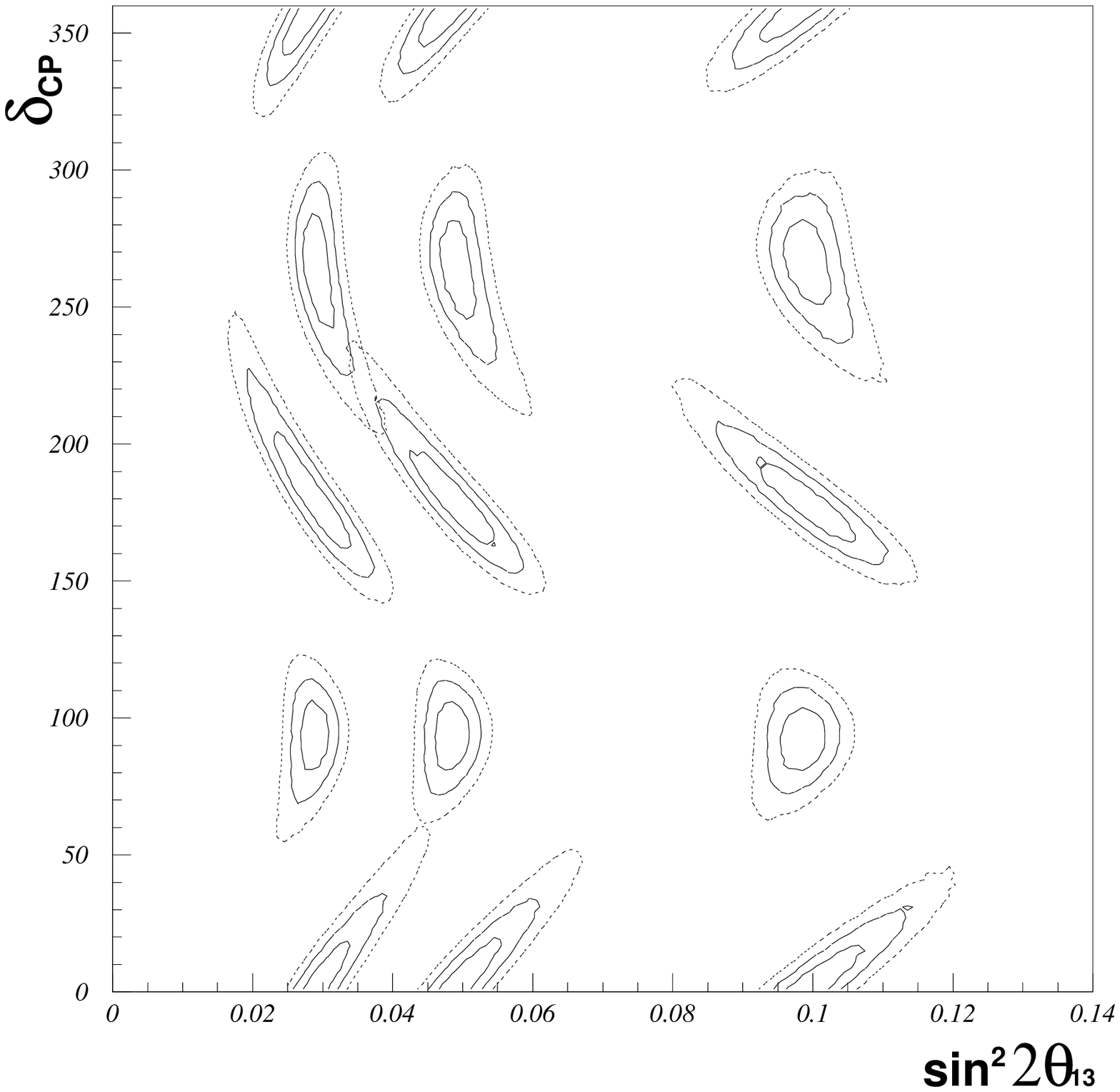}
\caption{\setlength{\baselineskip}{4mm}
 Allowed regions in the 100 MeV resolution (Gaussian sigma) case.
 Twelve allowed regions are overlaid for twelve true values,
 $\sin^2 2\theta_{13}$=0.1, 0.05, 0.02, and $\delta_{CP}$=0$^{\circ}$,
 90$^{\circ}$, 180$^{\circ}$, 270$^{\circ}$, respectively.
}
\label{Fig:Allowed_100mev}

\end{minipage}

\end{figure}

 As a reference, we first extract allowed regions in the perfect resolution case
  (See Figure~\ref{Fig:Allowed_0mev}).
 Twelve allowed regions are overlaid for twelve true values,
$\sin^2 2\theta_{13}$=0.1, 0.05, 0.02, and $\delta_{CP}$=0$^{\circ}$,
 90$^{\circ}$, 180$^{\circ}$, 270$^{\circ}$, respectively.
The $\delta_{CP}$ sensitivity is 20$\sim$30$^\circ$ depending 
on the true $\delta_{CP}$ value. 

\begin{figure}[htb]
\begin{center}
\includegraphics[angle=0, width=9.0cm]{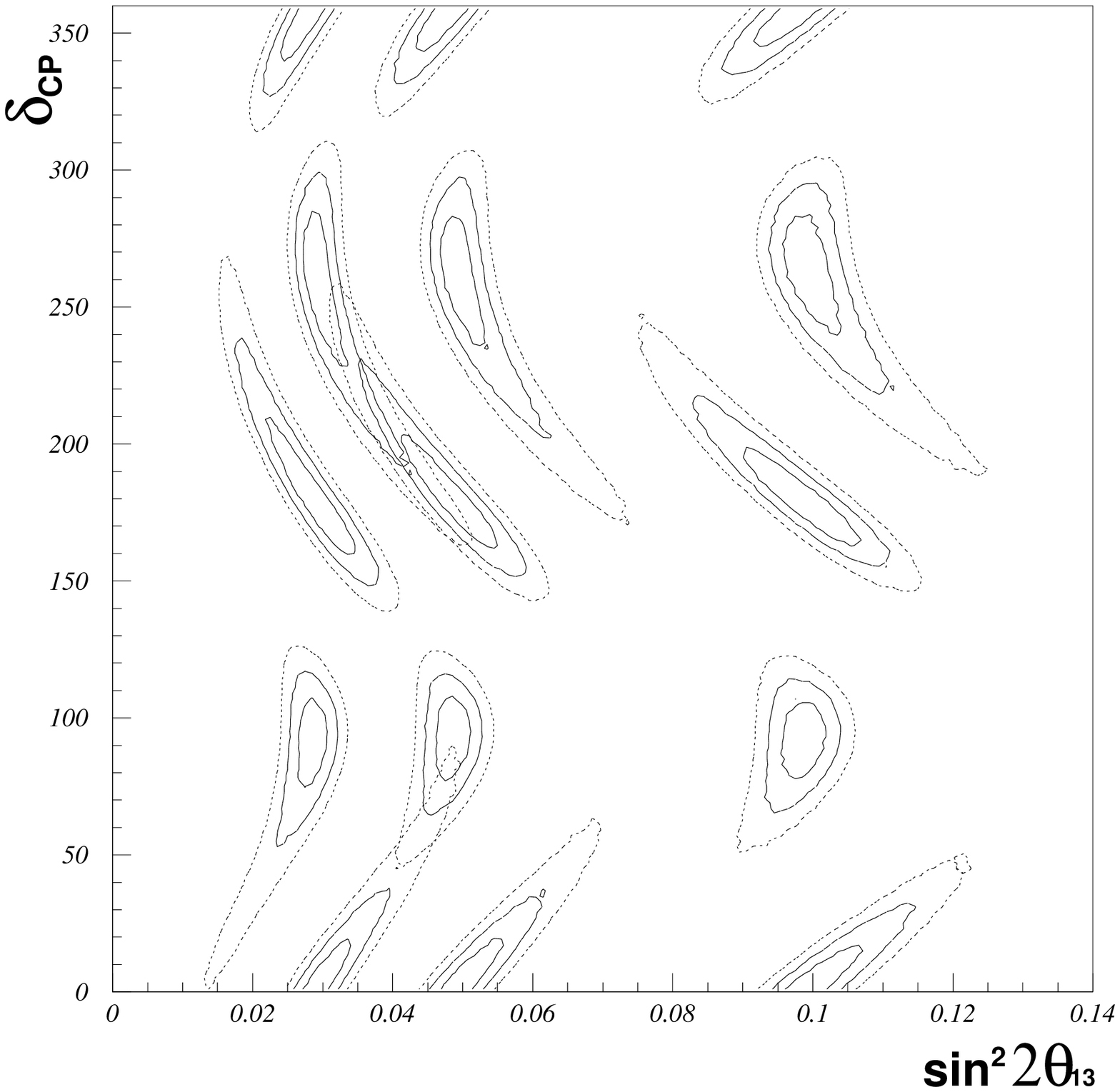}
\caption{\setlength{\baselineskip}{4mm}
 Allowed regions in the 200 MeV resolution (Gaussian sigma) case.
 Twelve allowed regions are overlaid for twelve true values,
 $\sin^2 2\theta_{13}$=0.1, 0.05, 0.02, and $\delta_{CP}$=0$^{\circ}$,
 90$^{\circ}$, 180$^{\circ}$, 270$^{\circ}$, respectively.
}
\label{Fig:Allowed_200mev}
\end{center}
\end{figure}

Allowed regions are then extracted for 100 and 200 MeV resolution cases.
The used energy spectra are same as Figures \ref{Fig:ES100mev} and 
\ref{Fig:ES200mev}. Results are shown as Figures \ref{Fig:Allowed_100mev} 
and \ref{Fig:Allowed_200mev}. 

One obvious but important issue to be pointed out is the robustness of the
fitting.The fit procedure shows that results could also be extracted with the 200 MeV 
resolution: this result is as expected statistically; however, we stress that in this case there is 
no second oscillation maximum peak visible in Figure~\ref{Fig:ES200mev}. Hence,
we think it is mandatory to keep an energy resolution less than 100 MeV as goal
for the credibility of this experiment.

\section{Investigation of mass hierarchy}
\label{sec:masshier}
\indent

The influence of matter on neutrino oscillations was first considered by Wolfenstein~\cite{Wolfenstein:1977ue}.  
As is well-known (although never directly experimentally verified), oscillation probabilities get modified under
these conditions. Matter effects are sensitive to the neutrino 
mass ordering and different for neutrinos 
and antineutrinos. As mentioned earlier, we consider in this paper the possibility
of a neutrino-only run. Hence, we briefly address in this section
the question of normal hierarchy (NH) versus inverted hierarchy (IH).
Since we are focusing on the potential discovery of CP-violation in the
leptonic sector, our discussion is geared towards possible ambiguities
that would arise if the mass hierarchy was unknown.

\begin{figure}[htbp]
\begin{center}
\includegraphics[angle=0, width=0.9\textwidth]{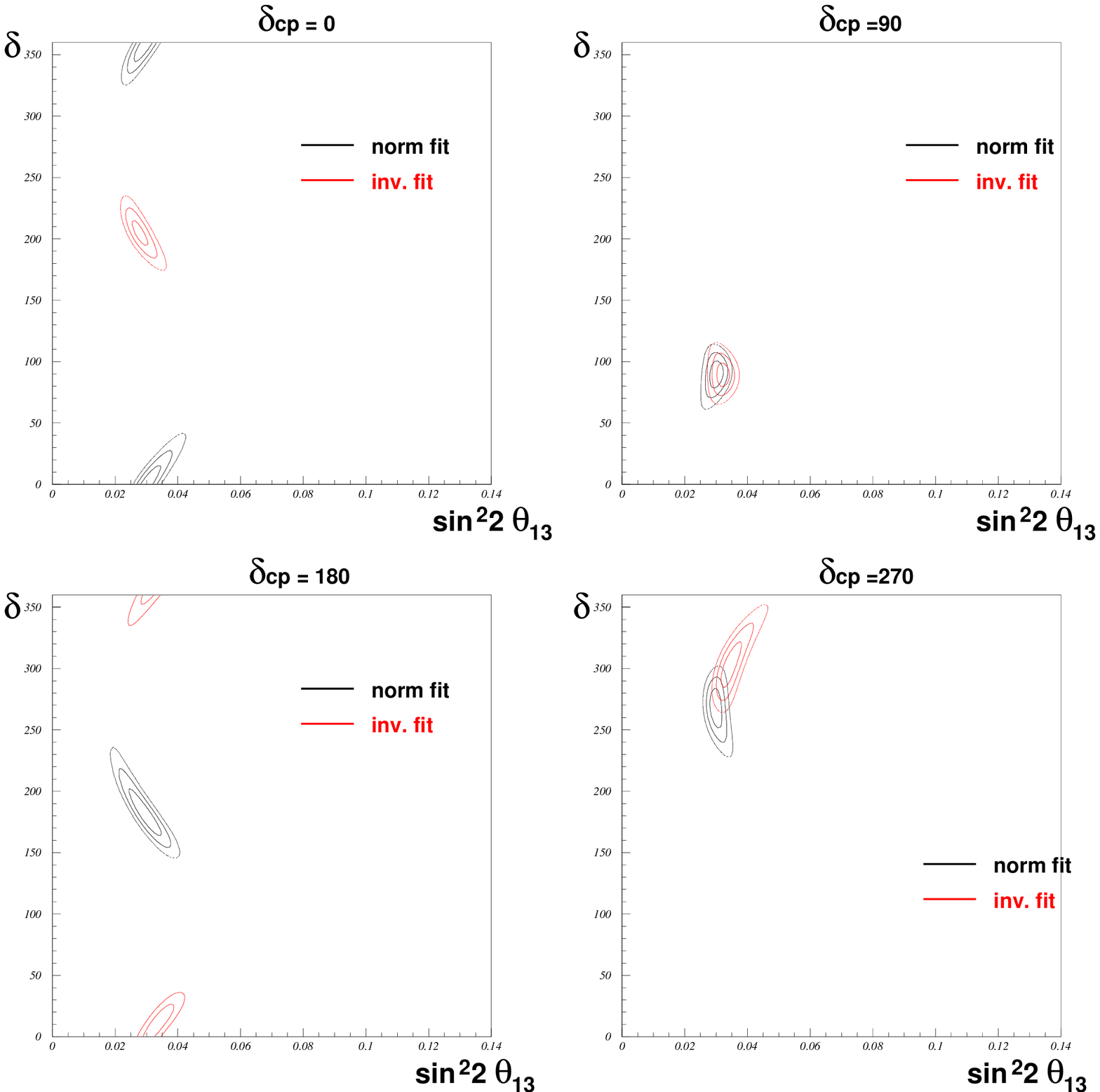}
\caption{\setlength{\baselineskip}{4mm}
 Mass hierarchy investigation with neutrino run only. If fits with both hierarchy hypotheses
 provide neither 0 nor 180$^\circ$, one can declare discovery of CP violation in the leptonic sector. If any of the fits results
in a $\delta_{CP}$ of 0 or 180$^\circ$, then an anti-neutrino run could be envisaged.
}
\label{Fig:DMresults}
\end{center}
\end{figure}

Indeed, Figure~\ref{Fig:DMresults} illustrates the results of a fit of a
pseudo-data with NH by both NH (black) and IH (red) hypotheses,
assuming only the neutrino run.
The best fit likelihood value with one assumed hierarchy is used to calculate the likelihood variation
$\Delta L$ for both hierarchies.  One could claim that CP-violation is discovered
if the experimental results of a given experiment exclude the $\delta_{CP}$ phase
to be either 0 or 180$^\circ$. Hence, the danger is that the lack of knowledge of the mass
hierarchy (or rather the ``wrong'' choice in hypothesis when selecting the hierarchy
in the fit of the data), gives a result  for $\delta_{CP}$ consistent with either 0 or 180$^\circ$.
On the other hand, if fits of a given experiment with both assumed hierarchies  
provide neither 0 nor 180$^\circ$, one can declare ``discovery'' although the mass hierarchy
could not be determined. Alternatively, if it turned out in the actual experiment
that one of results of the fit is consistent with $\delta_{CP}=0$ or 180$^\circ$, one would
still have the possibility to consider an anti-neutrino run in order to solve this
ambiguity.

\begin{figure}[htb]
\begin{center}
\includegraphics[angle=0, width=0.9\textwidth]{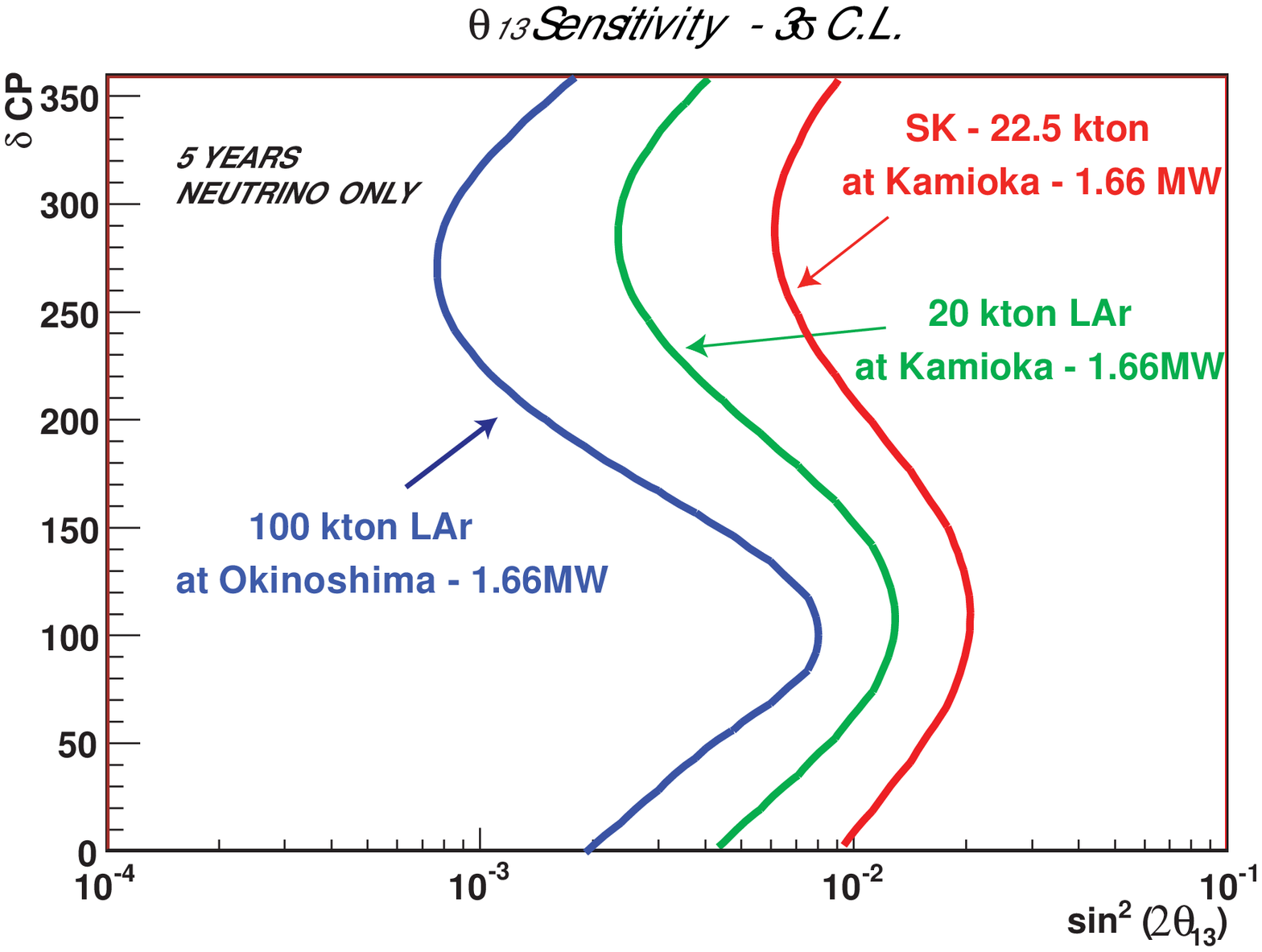}
\caption{\setlength{\baselineskip}{4mm}
 Sensitivity to $\sin^2 (2\theta_{13})$ for 5 years of neutrino run comparing
 100 kton LAr at Okinoshima with SK at Kamioka and 20 kton LAr at Kamioka. }
\label{Fig:t13sensi}
\end{center}
\end{figure}

Our results indicate that CP-violation would be unambiguously discovered under our assumption
for true values of $\delta_{CP}$'s in the region of 90$^\circ$ and 270$^\circ$. 
Alternatively, in case of  true values of $\delta_{CP}$'s near 0$^\circ$ or 180$^\circ$
an antineutrino flux would help untangle the two solutions if the neutrino mass
hierarchy was unknown.

\section{Expected sensitivity to $\sin^2 2\theta_{13}$}
\indent

Although this paper has focused on the possibility to measure the $\sin^22\theta_{13}$
and $\delta_{CP}$ oscillation parameters, it is instructive to estimate the sensitivity
of the potential setup in the case of negative result from T2K.
The corresponding sensitivity to discover $\theta_{13}$ in the true ($\sin^22\theta_{13},\delta_{CP})$
plane at 90\%~C.L. and $3\sigma$ is shown in Figure~\ref{Fig:t13sensi}, 
where we assumed  5 years of neutrino run comparing
 the 100 kton LAr at Okinoshima with increased statistics in SK
 at Kamioka and a  potential 20 kton LAr at Kamioka.

In order to discover a non-vanishing
$\sin^22\theta_{13}$, the hypothesis $\sin^22\theta_{13}\equiv 0$
must be excluded at the given C.L. As input, a true non-vanishing value
of $\sin^22\theta_{13}$ is chosen in the simulation and a fit with
$\sin^22\theta_{13}= 0$ is performed, yielding the ``discovery''
potential.
This procedure is repeated for every point in the ($\sin^22\theta_{13},\delta_{CP}$)
plane.

At the $3\sigma$~C.L. the sensitivity of the T2K experiment is 0.02~\cite{Itow:2001ee}. Continuing to
collect data at SK with an improved J-PARC neutrino beam for another 5~years 
would improve the T2K sensitivity by a factor $\sim 2$. In comparison,
our simulations indicate that a 20~kton LAr TPC is expected to perform
better than this although the masses are comparable, 
since the signal efficiency is higher than SK and the 
NC background is assumed to be negligible contrary to SK. Even better, a 100~kton LAr
TPC at Okinoshima would further improve the sensitivity by about a factor six compared
to SK at Kamioka (or a factor 10 compared to T2K), 
thanks to its bigger mass, increased cross-section at higher energies 
and reduced off-axis angle, although the neutrino fluxes at the same off-axis angle would
be reduced by a factor $\simeq 5$ because of the longer baseline.

\section{Neutrino energy resolution in a LAr medium}
\label{sec:eneres}
As mentioned previously, the neutrino energy resolution expected in LAr medium depends
on several parameters and their interplay that ultimately need to be measured experimentally.

In principle, energy and momentum conservation allow to estimate the incoming neutrino energy in the detector via a precise
measurement of decay products, with the exception of the smearing introduced by Fermi motion and 
other nuclear effects (nuclear potential, re-scattering, absorption, etc.) for interactions on bound
nucleons. In this section, we try to estimate the smearing introduced by these effects
with the help of exclusive neutrino event final states distributed with the Okinoshima flux and
generated with the GENIE MC~\cite{GENIEMC}  subsequently fully simulated with GEANT3.

Nuclear effects in neutrino interactions can be roughly divided into those of
the nuclear potential and those due to  reinteractions of decay products. 
Bound nucleons and other hadrons in nuclei are subject to a nuclear potential. The
Fermi energy (or momentum) must be calculated from the bottom of this nuclear
potential well, and the removal of a nucleon from any stable nucleus 
is  always an endothermic reaction. 
When hadrons are produced in the nucleus, some energy is spent to take it
out of this well: for a nucleon, the minimum energy is given by the nucleon separation
energy (around 8 MeV), and corresponds to a nucleon at the
Fermi surface. In this case, the daughter
nucleus is left on its ground state.  More deeply bound nucleons, 
leaving a hole in the Fermi sea, correspond to an excitation
energy of the daughter nucleus, and an additional loss of energy of the
final state products. This energy is then spent in evaporation and/or gamma
deexcitation. Thus, the energy of the final state products is
expected to be always slightly smaller than for interactions on free nucleons,
and spread over a range
of about 40 MeV.  Correspondingly, the Fermi momentum 
is transferred to the decay products and compensated by the recoil
of the daughter nucleus. Additional momentum distortions come from the 
curvature of particle trajectories in the nuclear potential.
 
  Reinteractions in the nuclear medium also play an important role. Interaction
products can lose part of their energy in collisions, or even be absorbed 
in the same nucleus where they have been created. This is particularly true
for pions, that have an important absorption cross section on nucleon
pairs, while kaons have smaller interaction probability.    

Once final state products have exited the nucleus, they will propagate
in the medium with the possibility that further interactions occur.

In the case of the liquid Argon TPC, the medium
is fully homogeneous and the detector is fully active. 
All deposited energy in the medium (above a certain threshold) will be eventually collected.
Several methods can be adopted to reconstruct the neutrino energy
and we list in the following three:

\begin{figure}[tbp]
\begin{center}
\includegraphics[angle=0, width=0.975\textwidth]{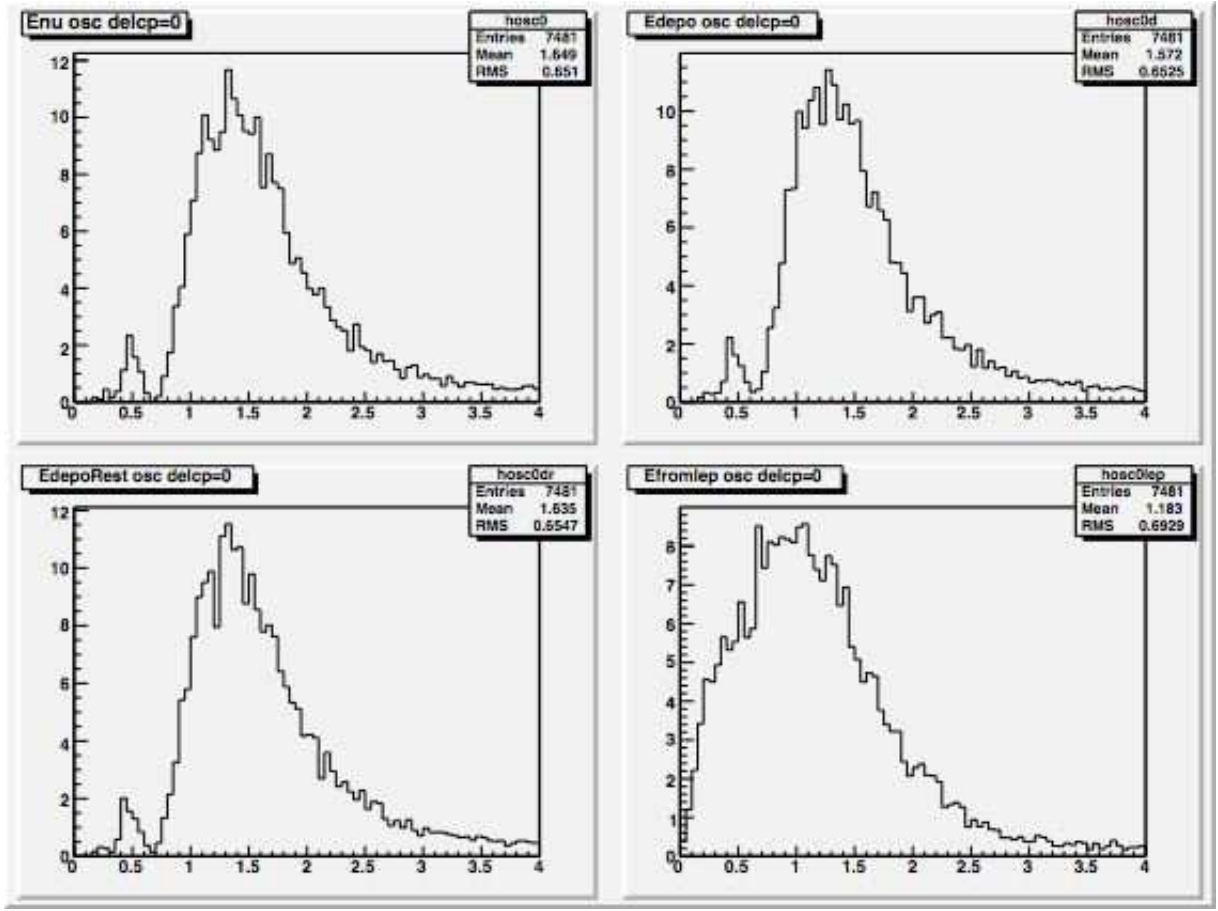}
\caption{\setlength{\baselineskip}{4mm}
Full GEANT simulations of the
reconstructed neutrino energy spectra from detailed event simulation (see text) for neutrino
oscillations with
 $\delta_{CP}$ = 0$^{\circ}$:
perfect reconstruction (top-left), 
LAr using deposited energy(right-top), 
ibid but with charged pion mass correction (left-bottom) 
using final state lepton only (like in WC) (right-bottom) cases. 
}
\label{Fig:ESgeantene}
\end{center}
\end{figure}

\begin{enumerate}
\item {\bf Final state lepton:} this method, traditionally used in large Water 
Cerenkov detectors like Kamiokande, IMB or SK, relies on the precise measurement of
the energy and direction of the outgoing lepton and kinematically
constrains the incoming neutrino energy by assuming a quasi-elastic
configuration: 
$$E_\nu = \frac{ME_\ell-m_\ell^2/2}{M-E_\ell+p_\ell\cos\theta_\ell}$$
where $E_\ell$, $p_\ell$ and $\cos\theta_\ell$ are the energy, momentum
and scattering angle of the outgoing lepton. 
This method is sensitive to the
final state configuration and to Fermi motion (up to $\simeq 240$~MeV/c)
which randomizes the direction of the outgoing lepton.

\item {\bf Momentum conservation:} neglecting the incoming neutrino mass,
one obtains
$$E_\nu = |\vec{P}_\nu| = |\vec{p}_\ell+\sum_h \vec{p}_h-\vec{P_F}|$$
where $\vec{p}_\ell$ is the momentum of the outgoing lepton,
$\vec{p}_h$ is the momentum of the outgoing hadron $h$ and
$\vec{P}_F$ is the Fermi motion of the hit nucleon (up to $\simeq 240$~MeV/c).
This method is also sensitive to 
Fermi motion since the recoiling remnant
hit nucleus (with momentum $-\vec{P}_F$) is not measured~\footnote{\setlength{\baselineskip}{4mm}We point out
that if one uses the knowledge on the direction of the incoming neutrino, the total (missing)
transverse momentum can be kinematically constrained to zero, thereby providing a handle
compensate for the transverse component of the Fermi motion. The longitudinal component
of the momentum, which is not negligible at low neutrino energies, remains however undetermined.}.

\item {\bf Energy conservation:} using a calorimetric approach, one can
sum the deposited energies of all outgoing particles. In a tracking-calorimeter
this is obtained by summing the $dE/dx$ measurements along each ionizing
track to obtain the associated kinetic energies $T\equiv \int (dE/dx) dx$.
One should identify final state particles in order to take into account their
rest masses. This method is sensitive to the nuclear binding energy of 
the decay products and is intrinsically more precise than the above method relying on
momenta for the energies considered here. Mathematically one can write this result as:
\begin{eqnarray}
E_\nu & = & E_{tot}-M = E_\ell+\sum_h E_h-M = E_\ell+\sum_h (T_h+m_h)-M \nonumber\\
& = & E_\ell+T_N+\sum_{\pi^\pm}\left( T_{\pi^\pm}+m_{\pi^\pm}\right) + \sum_\gamma E_\gamma+...
\end{eqnarray}
\end{enumerate}

All methods are sensitive to potential re-interactions of the outgoing hadrons within the
nucleus. In order to estimate the potential of a fully homogeneous and sensitive medium
like the liquid Argon TPC, we performed full simulations of neutrino interactions: the
event 4-vector are generated with the GENIE MC and final state particles (after nuclear
reinteraction) were propagated through the liquid Argon medium with GEANT3. 
The geometry of the setup is an infinite LAr box and at this stage we have neglected
charge quenching (i.e. we assume a linear response to $dE/dx$). We also assume
a 100\% efficiency to identify final state charged pions.

The achievable incoming neutrino energy resolution in the liquid Argon medium has
been estimated using $\nu_e$~CC events distributed with the $\nu_\mu$~flux
expected at the Okinoshima location. The results, using the energy conservation method,
are shown in Figure~\ref{Fig:ESgeantene}.

\section{Other baseline options}
\indent

 Although Okinoshima looks a very good candidate as mentioned so far,
the favorable geography would allow, in principle, for a few baseline candidates for the detector locations;
Kamioka for a relatively ''short'' baseline, Okinoshima 
for a medium baseline and Korea for the longest baseline.
It is worthwhile to compare the physics potential at Okinoshima with
other scenarios with the same strategy and same assumed
beam conditions, i.e. five years neutrino run at 1.66~MW.   
   
\begin{figure}[htbp]
\begin{center}
\includegraphics[angle=0, width=17.0cm]{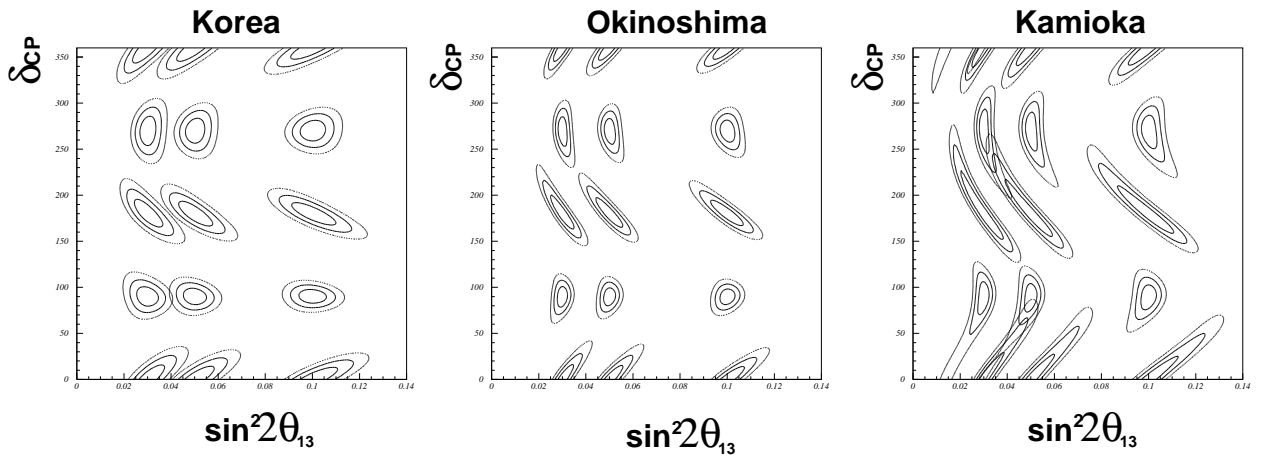}
\caption{\setlength{\baselineskip}{4mm}
 Allowed regions if the location of the 100 kton LAr TPC were at
 Kamioka, Okinoshima or Korea, assuming the T2K beam optics is kept.   
 Perfect resolution is assumed to illustrate the physics 
 potential from each site selection. 
}
\label{Fig:BLresults}
\end{center}
\end{figure}

Figure~\ref{Fig:BLresults} illustrates the analyzed allowed regions using each
 option. A perfect resolution was
assumed here to directly compare the physics potential
of the various sites. As seen, sensitivity of 
$\delta_{CP}$ is similar between Okinoshima and Korea, but that in Kamioka
is much worse than others. This indicates the analysis strategy using
only neutrino run is not suitable for Kamioka case, as was expected since
the OA2.5$^\circ$ available at Kamioka does not allow to cover 1st and 2nd
oscillation maxima peaks. 
As for the sensitivity of the $\sin^2 2\theta_{13}$,  Okinoshima is better than Korea,
due to the increased statistics at the shorter baseline.

\section{Proton Decay Discovery Potential}
Grand Unification of the strong, weak and electromagnetic interactions
into a single unified gauge group is an extremely appealing idea~\cite{Pati:1973uk,Georgi:1974sy}  which has
been vigorously pursued theoretically and experimentally for many years.
The detection of proton or bound-neutron decays would represent its
most direct experimental evidence.
The physics potentialities of very large underground Liquid Argon TPC
was recently carried out with detailed simulation of signal efficiency and
background sources, including atmospheric neutrinos and cosmogenic
backgrounds~\cite{Bueno:2007um}. It was found that a liquid Argon TPC,
offering good granularity and energy resolution, low particle detection threshold,
and excellent background discrimination, should  
yield  very good signal over background ratios in many possible
decay modes, allowing to reach partial lifetime sensitivities in
the range of $10^{34}-10^{35}$~years with exposures up to 1000~kton$\times$year,
 often in quasi-background-free conditions optimal for discoveries
 at the few events level, corresponding
to atmospheric neutrino background rejections of the order of $10^5$.
Multi-prong decay modes like e.g. $p\rightarrow \mu^- \pi^+ K^+$
or $p\rightarrow e^+\pi^+\pi^-$ and channels involving kaons like
e.g. $p\rightarrow K^+\bar\nu$, $p\rightarrow e^+K^0$ and $p\rightarrow \mu^+K^0$
are particularly suitable, since liquid
Argon imaging
provides typically an order of magnitude improvement in efficiencies for similar
or better background conditions compared to Water Cerenkov detectors.
Up to a factor 2 improvement in efficiency is expected for modes like $p\rightarrow e^+\gamma$
and $p\rightarrow \mu^+\gamma$ thanks to the clean photon identification
and separation from $\pi^0$. Channels like $p\rightarrow e^+\pi^0$ or $p\rightarrow \mu^+\pi^0$,
dominated by intrinsic nuclear effects,
yield similar efficiencies and backgrounds as in Water Cerenkov detectors. 
Thanks to the self-shielding and 3D-imaging properties of the liquid Argon TPC,
the result remains valid even at shallow depths where
cosmogenic background sources are important.
In conclusion, a LAr TPC would  not necessarily require very deep underground laboratories
even for high sensitivity proton decay searches. 

Once again, we stress the importance of an experimental verification of the physics potentialities
to detect, reconstruct and classify events in the relevant GeV~energy range.
This experimental verification will require the
collection of neutrino event samples with high statistics, 
accessible e.g. with a detector located at near sites of long
baseline artificial neutrino beams.

\section{Conclusions}

In this paper, we have reported our case study for a new future long baseline neutrino
and nucleon decay experiment with a 100~kton Liquid Argon TPC located in the region
of Okinoshima using the J-PARC Neutrino Facility.

Our study assumes a realistic upgrade of the J-PARC Main Ring operation yielding
an average 1.66~MW beam power for neutrinos coupled to the new far detector 
at Okinoshima, which would then see the beam at an off-axis angle of $\simeq 1^\circ$ and at a baseline of $\simeq 658$~km.

In this first discussion, we focused on the possibility to measure the $\theta_{13}$ and $\delta_{CP}$ parameters.
Our strategy is based on the precise measurement of the energy spectrum of the oscillated
events, and in particular on the comparison of the features of 1st maximum oscillation peak with those
of the 2nd maximum oscillation peak. In order to efficiently determine and study those features,
we rely on the excellent neutrino event reconstruction and incoming neutrino
energy resolution, as is expected in the case of the fully-sensitive and fine
charge imaging capabilities of the liquid Argon TPC.

Beyond this case study, we intend to perform more detailed simulations of the performance 
of the experiment including for example different sources of systematic errors and 
sources of backgrounds which have been up-to-now neglected.

The construction and operation of a 100~kton liquid Argon TPC certainly represents
a technological challenge at the present state of knowledge of the technique. 
For it to become a realistic option, we stress the 
importance and necessity of further R\&D and of dedicated experimental measurement
campaigns. At this stage, we intend to pursue our investigations on a ton-scale prototype 
based on the novel double-phase readout imaging method. Options to operate small
devices in the J-PARC neutrino beam are being assessed in parallel. In addition, 
we have started to address the possibility of an ``intermediate'' prototype, presumably
in the range of 1~kton of mass, based on a similar but scaled down design
of the potential 100~kton detector.

\section*{Acknowledgements}
Part of this work was supported by the Swiss National Foundation and ETH Zurich.


\end{document}